\documentclass[twocolumn,showpacs,preprintnumbers,amsmath,epsfig,epsf,amssymb]{revtex4}

\usepackage{graphicx}
\usepackage{dcolumn}
\usepackage{bm}

\begin{document}


\title{Black-hole ejecta by frame-dragging along the axis of rotation}

\author{Maurice H.P.M. van Putten}
 \affiliation{LIGO Laboratory, MIT 17-161, 175 Albany Street, Cambridge, MA 02139}

\date{\today}

\begin{abstract}
An energy $E=\omega J$ is derived for gravitational spin-orbit coupling 
by frame-dragging $\omega$ acting on angular momentum $J$. This interaction
defines a no-boundary mechanism for linear acceleration of magnetized matter 
along the axis of rotation of black holes. We explain GRB030329/SN2003dh by 
centered nucleation of a high-mass black hole with rapid rotation, producing 
baryon-poor ejecta by gravitational spin-orbit coupling. The duration of the
burst is attributed to spin-down in the emission of gravitational-waves 
by a surrounding non-axisymmetric torus.
\end{abstract}

\pacs{Valid PACS appear here}

\maketitle

GRB030329/SN2002dh showed that Type Ib/c supernovae are the parent populuation of
cosmological gamma-ray bursts. Type II and Type Ib/c supernovae are believed to 
represent core-collapse of massive stars \cite{fil97,tur03}. These events 
probably take place in binaries, such as in the Type II/Ib event SN1993J 
\cite{mau04}. This binary association suggests a hierarchy, in which hydrogen-rich, 
envelope rataining Type II events are associated with wide binaries, while 
hydrogen-poor, envelope stripped Type Ib and Type Ic are associated with 
increasingly compact binaries \cite{nom95,tur03}. By tidal coupling in the latter, 
the primary star rotates at the obital period. Consequently, core-collapse 
of the primary with an evolved core \cite{bro00} produce a rapidly rotating 
black hole \cite{lee02,bet03,van04b}. 

The Kerr metric \cite{ker63} establishes the exact solution of frame-dragging
induced black holes with nonzero angular momentum.
Frame-dragging appears in the form of nonzero angular velocities of zero angular 
momentum observers. It creates additional contributions to the
Riemann tensor \cite{cha83}. The Riemann tensor couples to spin \cite{pap51a,pap51b,pir56,mis74}, 
so that rotating black holes couple to the angular momentum of nearby particles. 
Indeed, specific angular momentum represents a rate of change of surface area and the
Riemann tensor is of dimension cm$^{-2}$ in geometrical units (Newton's constant $G=1$ 
and velocity of light $c=$1). Thus, curvature-spin coupling produces a force, whereby
test particles follow non-geodesic trajectories \cite{pir56}.

Consistent with the Rayleigh criterion -- the energy per unit of angular momentum is
small at large distances -- we find that rotating black holes couple to radiation as 
a channel to lower the total energy (of the black hole and radiation). 
With the second law of thermodynamics $dS\ge0$ on the entropy $S$, specific 
angular momentum increases with radiation:
\begin{eqnarray}
a_p \equiv \frac{-\delta J_H}{-\delta M} \ge \Omega_H^{-1} \ge 2M > M \ge a,
\label{EQN_RAYLEIGH}
\end{eqnarray}
based on the Kerr solution which has an angular velocity $\Omega_H\le 1/2M$,
where $M$ denotes the mass of the black hole with angular momentum $J_H$ and
specific angular momentum $a$. Nevertheless, 
isolated black holes are stable by exponential suppression due to 
due to canonical angular momentum barriers \cite{unr74,haw75,teu73,pre72,teu74}.

Here, we identify the potential energy induced by spin-orbit interactions. This
is a completely new result, which expresses a powerful interaction of 
frame-dragging along the axis of rotation of a black hole.
We begin with a new four-covariant derivation of curvature-spin coupling. This
result radically differs from the common view, that black-hole energetic processes
are confined to frame-dragging in the ergosphere.

\begin{figure*}
\includegraphics[scale=0.3]{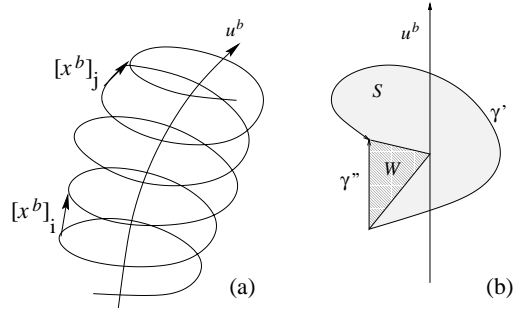}
\caption{
{\em Left:} 
Spacetime diagram of a particle in orbital motion. The orbital center and the 
orientation of the orbital plane are unconstraint. The separation vector 
$[x^b(t)]=x^b(t+T)-x^b(t)$ between successive orbits of period $T$ is carried 
along by parallel transport. It defines the tangent $u^b=[x^b]/T$ to the 
world-line of the orbital center, and its evolution.
{\em Right:} 
Curvature-spin coupling changes $u^b$ proportional to the surface area of $S$ 
and the wedge $W$ in a single orbit $\gamma^\prime$, closed by 
$\gamma^{\prime\prime}$.}
\label{FIG_51}
\end{figure*}

The world-line $x^a$ of a particle moving in a periodic orbit about the orbital 
center describes helical motion about the time-axis in spacetime. Fig. (\ref{FIG_51}) 
shows the closed curve $\gamma$ of a single orbit of period $T$ as measured in a 
local restframe, consisting of an open curve plus closing line-segment
\begin{eqnarray}
\gamma^\prime: x^b(t) ~(0<t<T),~~~\gamma^{\prime\prime}: t~(0<t<T).
\end{eqnarray}
The surface enclosed by $\gamma$ may be taken to be sum of the curved spiral surface 
$S$ and a closing wedge $W$
\begin{eqnarray}
S^{ab}=\int_{\gamma^\prime} x^{[a}v^{b]} ds,~~W^{ab}=Tw^{ab} = Tx^{[a}u^{b]},
\label{EQN_Q2}
\end{eqnarray}
where $\dot{}=d/ds$ and $v^b=dx^b/ds$ denotes the unit tangent to the particle 
world-line. This introduces the separation vector and four-velocity
\begin{eqnarray}
[x^b(t)]=x^b(t+T)-x^b(t),~~~u^b=\frac{[x^b(t)]}{T}
\label{EQN_SV}
\end{eqnarray}
of the particle between two consecutive orbits 

For parallel transport of a vector $\xi^b$ along $\gamma$ we have, 
according to the definition of the Riemann tensor \cite{lan84},
\begin{eqnarray}
\frac{\Delta \xi_{c}}{T} =\frac{1}{2T}R_{abcd}S^{ab}\xi^d
			           +\frac{1}{2}R_{abcd}w_n^{ab}\xi^d.
\label{EQN_4A}
\end{eqnarray}
Localizing to orbits of small radius $x^a$, the surface $S^{ab}$ is 
orthogonal to $u^b$. Consider, therefore,
\begin{eqnarray}
\dot{S}^{ab}=\frac{1}{T}\int_{\gamma^\prime} x^{[a}v^{b]}ds=
\epsilon^{ab}_{~~cd}s^cu^d
\end{eqnarray}
where $\dot{}=d/d\tau$ expressed in terms of the specific angular momentum $s^b$: 
a spatial vector orthogonal to $\dot{S}^{ab}$ whose magniture equals 
the rate of change of surface area. The variation satisfies
\begin{eqnarray}
\dot{\xi}_c=\frac{1}{2}\epsilon_{abef}R^{ef}_{~~cd}s^au^b\xi^d
            +\frac{1}{2}R_{abcd}w^{ab}\xi^d.
\label{EQN_QQ3}
\end{eqnarray}

In case of a point symmetric mass-distribution about the orbital center, such
as two particles attached to the end-points of a rod \cite{tho86}, a continuous
mass-distribution in a solid ring or charged particles in a magnetic field, we can 
integrate (\ref{EQN_QQ3}) over the mass-distribution. Since $w^{ab}$ is $2\pi$-periodic 
in the angular position of the wedge, only the term coupled to $s^b$ survives. Taking 
$\xi^b=u^b$, the acceleration of the orbital center satisfies
\begin{eqnarray}
\dot{u}_c=\frac{1}{2}\epsilon_{abef}R^{ef}_{~~cd}s^au^bu^d.
\label{EQN_QQ4}
\end{eqnarray}
This is curvature coupling to specific angular momentum \cite{pap51a,pap51b,pir56}. 
The particle trajectory becomes completely spefified by Fermi-Walker transport of $s^b$. 

The Kerr metric describes curvature induced by spin. This is the converse
of (\ref{EQN_QQ4}) above. Spinning bodies hereby couple to spinning bodies \cite{oco72}. 
Such interactions are commonly referred to as gravitomagnetic effects 
\cite{tho86} by analogy to magnetic moment-magnetic moment interactions, although
there is a difference in sign. To study this in the Kerr metric, we study
spin-orbit interactions along the axis of rotation. Based on dimensional 
analysis, the gravitational potential for spin aligned interactions should satisfy
\begin{eqnarray}
E=\omega J,
\label{EQN_USS}
\end{eqnarray}
where $\omega$ refers to the frame-dragging angular velocity produced
by the massive body and $J$ is the angular momentum of the spinning object.

The non-zero components of the Riemann tensor of the Kerr metric 
have been expressed in Boyer-Lindquist coordinates 
relative to tetrad 1-forms
${\bf e}_0=\alpha {\bf d}t,$
${\bf e}_1=\frac{\Sigma}{\rho}({\bf d}\phi-\omega{\bf d}t)\sin\theta,$
${\bf e}_2=\frac{\rho}{\sqrt{\Delta}}{\bf d}r,$ and
${\bf e}_3=\rho{\bf d}\theta$
by Chandrasekhar \cite{cha83}. Here, we use the expressions
\begin{eqnarray}
\Sigma^2=(r^2+a^2)^2-a^2\Delta \sin\theta,
~~\rho^2=r^2+a^2\cos^2\theta
\end{eqnarray}
and $\Delta = r^2 +a^2 -2Mr$ 
for a black hole of mass $M$ and specific angular momentum $a$.
In particular, one of the Riemann tensor 
component satisfies 
\begin{eqnarray}
R_{1302} = AD,
\end{eqnarray}
where $A ={aM}{\rho^{-6}}(3r^2-a^2\cos^2\theta)$ and $D 
=\Sigma^{-2}[2(r^2+a^2)^2+a^2\Delta\sin^2\theta].$ On-axis $(\theta=0)$, 
we have
\begin{eqnarray}
2A=-\partial_r\omega = \frac{2aM}{\rho^6}(3r^2-a^2),~~D=2.
\label{EQN_A10}
\end{eqnarray}
According to (\ref{EQN_QQ4}), this component of the Riemann tensor
creates the radial force
\begin{eqnarray}
F_2=JR_{3120}=JAD=-\partial_2\omega J.
\label{EQN_QQ6}
\end{eqnarray}
The assertion (\ref{EQN_USS}) follows from
\begin{eqnarray}
E=\int_r^\infty F_2 ds = \omega J.
\label{EQN_USS1}
\end{eqnarray}

The result (\ref{EQN_USS1}) also follows from a completely independent derivation, 
by considering the
difference in total energy between two particles in counter rotating orbits about 
the axis of rotation of the black hole. Let $u^b$ denote the velocity four-vector 
and $u^\phi/u^t=\Omega$ the angular velocities of either one of these,
$-1=u^cu_c=[g_{tt}+g_{\phi\phi}\Omega(\Omega-2\omega)](u^t)^2.$
This normalization condition has the two roots
\begin{eqnarray}
\Omega_\pm=\omega \pm \sqrt{\omega^2 - (g_{tt}+(u^t)^{-2})/g_{\phi\phi}}.
\end{eqnarray}
We insist that these two particles have angular momenta of opposite sign and 
equal magnitude, $J_\pm = g_{\phi\phi} u^t (\Omega_\pm - \omega)$,
\begin{eqnarray}
J_\pm =g_{\phi\phi}u^t 
 \sqrt{\omega^2 - (g_{tt}+(u^t)^{-2})/g_{\phi\phi}}=\pm J.
\label{EQN_EJ}
\end{eqnarray}
This shows that $u^t$ is the same for each particle. The total
energy of the particles is given by $E_\pm = (u^t)^{-1}+\Omega_\pm J_\pm,$
and hence one-half their difference
\begin{eqnarray}
E=\frac{1}{2}(E_+-E_-)=\omega J.
\label{EQN_USS2}
\end{eqnarray}

The above shows that curvature-angular momentum coupling (\ref{EQN_USS1}) is 
universal: it applies whether the angular momentum is mechanical, 
electromagnetic or quantum mechanical in origin. 

\begin{figure}
\includegraphics[scale=0.3]{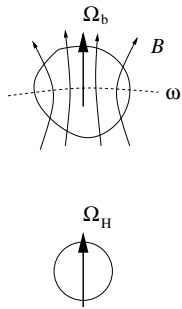}
\caption{Gravitational spin-orbit coupling betweem frame-dragging $\omega$ and
the angular momentum $J$ of the blob produces a potential energy $E=\omega J$. 
In the limit of perfect conductivity, the blob assumes a uniform angular 
velocity $\Omega_b$ with $J\propto \Omega_b-\omega$. The blob is ejected 
when $E>0$ and absorbed if $E<0$. 
This spin-orbit interaction is a no-boundary mechanism for linear acceleration: 
it does not require any contact with the event horizon of the black hole for the 
formation of baryon-poor ejecta.}
\label{FIG_52}
\end{figure}

A magnetized blob of perfectly conducting fluid is characterized by rigid
rotation \cite{tho86} with angular velocity $\Omega_b$. In the frame of
zero angular momentum observers, the local charge-density is given by the
Golreich-Julian charge density \cite{gol69}. In their
lowest energy state characterized by vanishing canonical angular momentum,
their angular momentum satisfies $J=eA_\phi$, where $e$ denotes the
unit of electric charge and $A_\phi$ the $\phi-$component of the 
electromagnetic vector potential $A_a$ \cite{van00}. Consider a magnetized blob 
about the axis of rotation of the black hole shown in Fig. 
(\ref{FIG_52}). The number density $N(s)$ of particles
per unit distance $s$ along the axis of rotation -- the number density per
unit scale-height for a given magnetic flux -- satisfies
\begin{eqnarray}
N(s)=\frac{\rho}{e}=(\Omega_b-\omega)A_\phi.
\end{eqnarray}
A pair of blobs in both directions along the spin-axis of scale height $h$ 
hereby receives an energy
\begin{eqnarray}
E_{blob}= \omega JNh = \omega(\Omega_b-\omega)A^2_\phi h.
\label{EQN_EBLOB}
\end{eqnarray}
Expressed in dimensionful form, we have
\begin{eqnarray}
E_{blob} = \left(1\times 10^{47}\mbox{erg}\right)
           B_{15} h_M^3 H,
\end{eqnarray}
where $h_M=h/M$ denotes the linear dimension of the blob, $B_{15}=B/10^{15}$.
Here, we use the normalized function
$H=4\hat{\omega}\left(\hat{\Omega}_b-\hat{\omega}\right)$,
where $\hat{\omega}=\omega/\Omega_H$ and $\hat{\Omega}_b=\Omega_b/\Omega_H$.

We propose that blobs are ejected along the axis of rotation {\em both} to 
infinity ($H>0$) {\em and} into the black hole ($H<0$). 
This happens, for example, by break-up of a blob of perfectly conducting 
fluid with vanishing total angular momentum into two peaces, by tidal
interaction of gravitational spin-orbit coupling given that $\omega$
is strong near the black hole and weak at larger distances.

A discrete, intermittent process produces ``pancakes," which produce 
gamma-rays by shocks when they collide (see \cite{pir04} for a review). 
This forms an alternative to continuous outflows in the form of baryon-poor 
jets (see \cite{lev04} for a recent account), wherein shocks appear either
by steepening due to temporal fluctuations \cite{lev97} or in the interaction 
with the environment.

We explain GRB-supernova GRB030329/SN2003dh in terms of core-collapse
supernova \cite{woo93,kat94} by centered nucleation of a rapidly
rotating black hole in a massive star, followed by the ejection of baryon-poor
blobs and radiative spin-down of the black hole against gravitational radiation 
catalyzed by a surrounding non-axisymmetric torus \cite{van01}. We attribute the
supernova to irradiation of the remnant stellar envelope from within, by
high-energy photons produced in the dissipation of a subdominant torus wind.

{\bf Acknowledgment.}
The author thanks A. Levinson, E. Schuryak and R.P. Kerr for constructive comments. This research is
supported by the LIGO Observatories, constructed by Caltech and MIT
with funding from NSF under cooperative agreement PHY 9210038.
The LIGO Laboratory operates under cooperative agreement
PHY-0107417. This paper has been assigned LIGO document number
LIGO-P0400XX-00-R.

\end{document}